%% file: pppmbs.tex
\documentclass[10pt,twocolumn]{IEEEtran}

\usepackage[scaled=0.85]{DejaVuSansMono}
\usepackage[dvipsnames,svgnames]{xcolor}
\usepackage{csquotes}
\usepackage{xspace}
\usepackage{amsmath}
\usepackage{subcaption}
\usepackage{multirow}
\usepackage{graphicx}
\usepackage{listings}
\usepackage[svgnames]{xcolor}
\usepackage{mathrsfs}
\usepackage{mathtools}
\usepackage{phoenician}
\usepackage{cite}

\definecolor{cinnamon}{rgb}{0.82, 0.41, 0.12}
\definecolor{ColorForLink}{named}{DarkRed}
\definecolor{ColorForCite}{named}{DarkOliveGreen}
\definecolor{URLLinkColor}{named}{MidnightBlue}

\usepackage[pagebackref=true,breaklinks=true,letterpaper=true,colorlinks=true,bookmarksnumbered=true,bookmarksopen=true,bookmarksopenlevel=1]{hyperref}

\hypersetup{
    pdftitle={A Metric for Performance Portability},
    pdfauthor={S.J. Pennycook, J.D. Sewall and V.W. Lee},
    pdfsubject={We propose language and a metric for discussions of performance portability.},
    pdfkeywords={portability, programming models, performance measurement},
    linkcolor=ColorForLink,
    citecolor=ColorForCite,
    urlcolor=URLLinkColor
  }

\usepackage{paralist}

\newcommand{\eg}{\textit{e.g.}\xspace}
\newcommand{\ie}{\textit{i.e.}\xspace}

\newcommand{\paper}{paper\xspace}

\newcommand{\pp}{\;\mathclap{\reflectbox{$\mathrm{P}$}}\mathrm{P}}
\newcommand{\ppm}{$\pp$\xspace}

\begin{document}
\renewcommand{\thelstlisting}{\alph{lstlisting}}

\title{A Metric for Performance Portability}
\author{S. J. Pennycook, J. D. Sewall and V. W. Lee\\
Intel Corporation\\
Santa Clara, California\\
\{john.pennycook,jason.sewall,victor.w.lee\}@intel.com}

\maketitle

\input{abstract.tex}
\input{introduction.tex}
\input{definition.tex}
\input{measurements.tex}
\input{discussion}
\input{summary.tex}

\newpage
\section{Disclaimers}\label{sec:disclaimers}

{\small
\noindent Intel, the Intel logo, Intel Xeon, Intel Xeon Phi and Intel VTune are trademarks of Intel Corporation or its subsidiaries in the U.S. and/or other countries.

Software and workloads used in performance tests may have been optimized for performance only on Intel microprocessors. Performance tests, such as SYSmark and MobileMark, are measured using specific computer systems, components, software, operations and functions. Any change to any of those factors may cause the results to vary. You should consult other information and performance tests to assist you in fully evaluating your contemplated purchases, including the performance of that product when combined with other products.  For more complete information visit  www.intel.com/benchmarks.

Intel does not control or audit third-party benchmark data or the other papers referenced in this document. You should visit the referenced documents and confirm whether referenced data are accurate.
}

\section*{Acknowledgements}
{\small
\noindent The authors wish to thank: the rest of the HEAT team -- Alex Duran, Jeongnim Kim, Amrita Mathuriya, Larry Meadows, Roland Schulz and Dayle Smith -- for feedback on the draft of this \paper; Tom Deakin and Simon McIntosh-Smith at the University of Bristol, for comments on our metric; and the attendees of the Department of Energy Centres of Excellence Performance Portability meeting in April 2016, for interesting discussions that seeded this work.
}

\bibliographystyle{abbrv}
\bibliography{performance_portability}
\end{document}

%% file: abstract.tex
\begin{abstract}
The term ``performance portability'' has been informally used in computing to refer to a variety of notions which generally include: 1) the ability to run one application across multiple hardware platforms; and 2) achieving some notional level of performance on these platforms.  However, there has been a noticeable lack of consensus on the precise meaning of the term, and authors' conclusions regarding their success (or failure) to achieve performance portability have thus been subjective.  Comparing one approach to performance portability with another has generally been marked with vague claims and verbose, qualitative explanation of the comparison.  This \paper presents a concise definition for performance portability, along with a simple metric that accurately captures the performance and portability of an application across different platforms. The utility of this metric is then demonstrated with a retroactive application to previous work.
\end{abstract}
%

%% file: introduction.tex
\section{Introduction}


\noindent Developing and maintaining a separate implementation of an application for each new hardware platform of interest is a huge undertaking, and one that is unrealistic for the developers of the large and long-lived applications found in high performance computing (HPC).  It is therefore unsurprising that the introduction of programming models, languages and tools that provide developers the ability to write single-source codes that can be compiled for different targets (\eg OpenCL\footnote{\label{foot:tm}Other names and brands may be claimed as the property of others.}~\cite{OpenCL}, OpenMP\footnotemark[\getrefnumber{foot:tm}]~\cite{OpenMP}, OpenACC\footnotemark[\getrefnumber{foot:tm}]~\cite{OpenACC}) has been met with enthusiasm from the HPC community.

However, the growth of options for ensuring \emph{functional portability} across diverse hardware platforms naturally leads one to wonder if any of them can deliver portability without sacrificing the level of performance that developers have come to expect from traditional HPC languages (such as C, C++ and Fortran).  The term \emph{performance portability} has appeared in the literature, but the community has not agreed upon the meaning of the term nor how to measure the degree to which an application (or library, framework, programming model, etc) has become performance portable~\cite{DoE-PP-Meeting}.

A shared lexicon and shared goals are necessary first steps in bringing performance portability researchers together to discuss and compare their findings, and to that end this \paper makes the following contributions:
\begin{enumerate}
\item We propose a new definition of performance portability, and demonstrate how it avoids the shortcomings of previous definitions proposed in the literature;
\item We describe a novel metric for characterizing performance portability (according to our proposed definition) and demonstrate its accuracy and utility for quantifying an application's performance \emph{and} portability; and
\item We retroactively apply our metric to a number of published application studies, thereby highlighting the utility of a shared metric when comparing and contrasting different approaches to performance portability.
\end{enumerate}

\section{Related Work}

\noindent There have been a number of efforts to develop new programming models, languages and tools that provide users with a productive means of achieving performance portability.  Some have proposed the use of domain-specific languages (DSLs), providing a limited set of high-level abstractions for a specific domain alongside a mechanism for generating optimized platform-specific binaries (\eg Liszt~\cite{Liszt}, OP2~\cite{OP2,PyOP2,OPS}, STELLA~\cite{STELLA}, Pochoir~\cite{Pochoir}, PATUS~\cite{PATUS}, Halide~\cite{Halide} and Nabla~\cite{Nabla}).  Others have proposed portability frameworks that take a different approach, providing a lower-level abstraction of performance-enabling hardware for developers to program against: some of these tools are truly languages or language extensions (\eg OpenCL~\cite{OpenCL}, OpenMP 4.0~\cite{OpenMP}, OpenACC~\cite{OpenACC}, Petabricks~\cite{Petabricks} and Sequoia~\cite{Sequoia}) while others are provided as libraries (\eg Kokkos~\cite{Kokkos} and RAJA~\cite{RAJA}).

The performance portability claims of these languages and frameworks have subsequently been tested in several individual application case-studies~\cite{Pennycook-MD,Pennycook-OpenCL,Deakin-GPUSTREAM,SMS-Grid,OPS,Sabne-OpenACC,Zhang-OpenCL,Herdman-LOC,Rul-OpenCL,Komatsu-evaluating,Larkin-PP}.  The amount of expended development effort focused on platform-specific optimization and the level of performance that an application must achieve in order to be considered performance portable is inconsistent across studies; we explore these inconsistencies and examine the results of some of these studies in more detail later in the \paper.

%

%% file: definition.tex
\section{Defining Performance Portability}
\noindent As discussed, there is far from consensus on the meaning of \emph{performance portability}, and one goal of this \paper is to establish a significant and useful definition for the term.  To begin with, we establish some basic terms: a \emph{platform} is a particular execution environment (\ie hardware, an operating system, some compilation and runtime tools); and an \emph{application} is any suite of software that can accept a given \emph{problem} as input and produce an output that can be validated against some existing measure of correctness (\ie two completely separate source codes that solve the same problem to satisfiable accuracy are the same application).

These are each inclusive concepts that primarily delineate components needed for further discussion.  We also establish the following definitions for performance and portability:

\blockquote{\textbf{Performance}\\Any measurable property of an application's correct execution of a problem on a platform.}

\noindent This is a broad definition; the most common performance metrics are based on economy of time (\eg time-to-solution or floating-point operations/second) or economy of energy (\eg energy-to-solution or floating-point operations/watt). Also note that it is rare to be able to make a meaningful comparison between performance measurements based on distinct properties.

\blockquote{\textbf{Portability}\\The ability of an application to execute a problem correctly on a given set of platforms.}

\noindent An application and problem combination is either portable for a given set of platforms or not; it either runs correctly on all of them or fails for at least one of them.  This is distinct from the portability of an application's source code.

\subsection{Criteria for Useful Definitions of Performance Portability}

\noindent There is a set of criteria that a useful definition of performance portability should satisfy; the definition should:
\begin{enumerate}
\item Reflect the individual meanings of the terms ``performance'' and ``portability'';
\item Be objective; and
\item Be measurable (and comparisons of the measured value should be meaningful).
\end{enumerate}

\noindent Satisfying these criteria ensures: that usage of the term is in line with a reader's expectations (\ie that performance portable applications are efficient on multiple platforms); that claims of performance portability are backed by facts and not subject to opinions of programming languages and platforms; that there is a standard way to evaluate the success of the community efforts to achieve performance portability; and that an end-user given a choice of two applications is able to reason about which is the most performance portable across the hardware platforms that are available to them.

Although \emph{productivity} is one of the driving factors behind community interest in solutions to performance portability~\cite{DoE-PP-Meeting}, we deliberately exclude it from our definition criteria because it is at odds with objectivity and measurability -- an application developer's productivity heavily depends upon their skillset, and common productivity metrics (\eg lines/words of code) do not reflect the reality that a library/framework may be more productive to use than it is to develop and/or maintain.  Whether or not there exists a meaningful and useful definition for \emph{productive} performance portability is beyond the scope of this work, but keeping productivity considerations independent of performance portability is compatible with existing efforts to measure the impact of different performance portability frameworks on programmer productivity~\cite{Herdman-LOC,Martineau-TeaLeaf,Sandia-Codesign-Milestone}.

\subsection{Existing Definitions}

\noindent In this section, we test a number of existing definitions for performance portability against our criteria.  While we respect the intentions behind these definitions and agree with aspects of all of them, they each have shortcomings and would benefit from further refinement.  This is by no means an exhaustive list of definitions -- rather, we have selected a set of representative perspectives from both literature and informal discussions at conferences and translated them into definitions.

\blockquote{\textbf{Definition 1}\\An approach to application development, in which developers focus on providing portability between platforms without sacrificing performance.} 
This definition serves as an important reminder: if a developer is working to support multiple platforms and to improve performance across the platforms they support, then they should be able to claim some level of performance portability.  While we agree with this aspect of the definition, it is highly subjective and sets a very low bar.

\blockquote{\textbf{Definition 2}~\cite{Larkin-PP}\\The ability of the same source code to run productively on a variety of different architectures.}
This definition implies that maintaining separate code paths for each platform is not productive, and should therefore be disallowed. However, the number of device-specific optimizations and other source code changes (\eg additional pragmas) that are permitted before a code is no longer ``the same'' as it was originally is subjective.

\blockquote{\textbf{Definition 3}~\cite{Zhu-Earth}\\${P_n}^{\mathscr{P}}(b \rightarrow t) = \frac{{S_n}^t}{{S_n}^b}\times100\%$ for program $\mathscr{P}$, base system $b$, target system $t$ and speed-up on $n$ nodes $S_n$.}
This definition is clearly objective and measurable, but comparison of these measurements is complicated (and may not be meaningful) because the scaling behavior of two applications does not necessarily reflect their absolute performance.

\blockquote{\textbf{Definition 4}~\cite{SMS-Grid}\\The ability of an application to achieve a similar high fraction of peak performance across target devices.}
``Peak performance'' here is the theoretical maximum performance of a platform, and the fraction of this that an application can achieve is an objective and useful quantity that we refer to as \emph{architectural efficiency}.  This definition is measurable and comparable, but its use of the terms ``similar'' and ``high'' make it subjective.  Furthermore, the aspiration that similar architectual efficiencies be achieved across platforms is often unrealistic and penalizes sets of platforms with microarchitectural differences (\eg applications with an imbalanced ratio of multiplications and additions are more efficient on platforms without fused multiply-add instructions).

\blockquote{\textbf{Definition 5}~\cite{Kokkos}\\The ability of an application to obtain the same (or nearly the same) performance as a variant of the code that is written specifically for that device.}
This definition meets all of our criteria except for objectivity and measurability: ``nearly the same'' is subjective, and different development teams are likely to have different opinions on how similar performance must be before they have achieved performance portability.

\blockquote{\textbf{Definition 6}\\The ability of an application to execute with a performance difference of less than 2$\times$ on two different systems, without significant software changes.}
This definition initially appears to satisfy all of our criteria, but it has counterintuitive properties that do not reflect absolute performance.  For example, consider an application that executes on Platform A in four seconds and on Platform B in one second; a speed-up of 2$\times$ on Platform A \emph{or} a slow-down of 2$\times$ on Platform B both result in a state that satisfies the definition.  Like Definition 4, it also fails to account for architectural differences between platforms that may prevent an application from meeting the definition.

\subsection{Our Definition}\label{sec:pp-definition}

\noindent Our proposed definition of performance portability is a refinement of the definitions above:

\blockquote{\textbf{Performance Portability}\\A measurement of an application's performance efficiency for a given problem that can be executed correctly on all platforms in a given set.}

\noindent An application that cannot execute a given problem correctly across a given set of platforms is not performance portable (for that problem across that set of platforms).  We separate application and problem (\ie input parameters) to acknowledge that an application's behavior and performance is largely dependent on the problem being solved, often to the point that a different algorithm may be more appropriate.

Our definition builds directly on our earlier definitions of performance and portability, and remains objective by specifying that performance efficiency (\ie a ratio of observed performance relative to some proven, achievable level of performance such as a previous architectural study or a performance model) for each platform be used, rather than absolute performance. This precludes the use of terms like ``good performance''.  It is not obvious from the definition alone that it satisfies the third of our criteria; the next section presents a new metric for performance portability to support our proposed definition.


%% file: measurements.tex
\section{Measuring Performance Portability}

\noindent As with our definition, we have designed our performance portability metric around specific criteria.  A useful metric should:
\begin{enumerate}
 \item Be measured specific to a set of platforms of interest $H$.
 \item Be independent of the absolute performance across $H$.
 \item Be zero if a platform in $H$ is unsupported, and approach zero as the performance of platforms in $H$ approach zero.
 \item Increase if performance increases on any platform in $H$.
 \item Be directly proportional to the sum of scores across $H$.
\end{enumerate}

\noindent Satisfying these criteria ensures that the metric is easy to understand: that there are no assumptions about the number of or type of platforms; that an application cannot claim performance portability in situations that it couldn't claim portability; and that comparisons of performance portability reflect comparisons of performance.

Our proposed metric is the harmonic mean of an application's performance efficiency observed across a set of platforms.  If the application fails on any measured platform(s), then we define the performance portability to be 0.

Formally, for a given set of platforms $H$, the performance portability \ppm of an application $a$ solving problem $p$ is:
\[
 \pp(a,p,H) =
 \begin{cases}
  \dfrac{|H|}{\sum_{i \in H} \dfrac{1}{e_i(a,p)}} & \text{if } i \text{ is supported } \forall i \in H \\
  0                                             & \text{otherwise} \\
 \end{cases}
\]

\noindent where $e_i(a,p)$ is the performance efficiency of application $a$ solving problem $p$ on platform $i$.

The harmonic mean has been previously demonstrated as a useful way to aggregate multiple performance numbers~\cite{Smith88}.  Unlike the geometric and arithmetic means, the harmonic mean satisfies criteria 3) and 5) above.  It is also robust to large outliers, preventing applications from making \ppm artificially large by adding more platforms; this ensures that, given measurements for two applications on the same set of platforms, it is clear which application is most likely to achieve the best performance on any selected platform.

There are multiple performance efficiency metrics that could be used to compute \ppm. In this \paper, we consider two: 1) \emph{architectural efficiency} (achieved performance as a fraction of peak theoretical hardware performance), which represents the ability of an application to utilize hardware efficiently; and 2) \emph{application efficiency} (achieved performance as a fraction of best observed performance)~\cite{Satish-Ninja}, which represents the ability of an application to use the most appropriate implementation and algorithm for each platform.  This distinction is important, because an application that hits 100\% of peak performance (\eg DRAM bandwidth or GFLOP/s) is not necessarily well-optimized (\eg a bandwidth-bound application may be moving data between DRAM and cache unnecessarily, or a compute-bound application may be performing redundant computation), and naturally the best known performance of an application is not necessarily the best that is possible.  In both cases, performance efficiency can be represented as the ratio of observed and ceiling (\ie peak theoretical or best observed) performance arranged such that it lies in $[0,1]$\footnote{Or an equivalent percentage in $[0,100]$}; 1 means that the observed performance matches the best, and lower values show how much worse the observed is than the best.

\subsection{Constructed Examples}

\noindent Table~\ref{tbl:example}\subref{tbl:example-eff} contains some example performance data from an imaginary application running on five different imaginary platforms.  The example has been constructed to represent the sorts of complications that may arise when comparing performance across different hardware platforms: the implementation used for Platform B trades communication for redundant computation, performing twice as many floating-point operations as the other platforms; Platforms A and C are microarchitecturally very similar but for the introduction of fused multiply-add instructions in Platform C; and Platform D is a new platform that the application doesn't yet support.

\begin{table}[t!]
 \small
 \centering
 \begin{subtable}[t]{\linewidth}
  \setlength{\tabcolsep}{4pt}
  \centering
  \begin{tabular}{c||c|c|c||c|c|c}
  \multirow{2}{*}{\textbf{Platform}} & \multicolumn{2}{c|}{\textbf{GFLOP/s}} & \textbf{Arch} & \multicolumn{2}{c|}{\textbf{Time (s)}} & \textbf{App.} \\
  & \textbf{Achieved} & \textbf{Peak} & \textbf{Eff.} & \textbf{Achieved} & \textbf{Best} & \textbf{Eff.} \\ \hline
  A &  40 &  250 & 16\% & 25.0 & 25.0 & 100\% \\ \hline
  B & 160 &  700 & 23\% & 12.5 & 10.0 &  80\% \\ \hline
  C &  40 &  500 &  8\% & 25.0 & 12.5 &  50\% \\ \hline
  D &   - &  800 &  0\% &    - &  6.0 &   0\% \\ \hline
  E & 200 & 1250 & 16\% &  5.0 &    1 &  20\% \\
  \end{tabular}
  \caption{Application and architectural efficiencies.\medskip}\label{tbl:example-eff}
 \end{subtable}
 \begin{subtable}[t]{\linewidth}
  \centering
  \begin{tabular}{c|c|c}
  \multirow{2}{*}{\textbf{Platform Set ($H$)}} & \multicolumn{2}{c}{\textbf{\boldmath$\pp(a,p,H)$}} \\
                                      & \textbf{Arch. Eff.} & \textbf{App. Eff.} \\ \hline
  \{A, B, C, D, E\} &  0.00\% &   0.00\% \\ \hline
     \{A, B, C, E\} & 13.62\% &  43.23\% \\ \hline
        \{A, B, C\} & 12.97\% &  70.59\% \\ \hline
           \{A, C\} & 10.67\% &  66.67\% \\ \hline
              \{A\} & 16.00\% & 100.00\% \\
  \end{tabular}
  \caption{Performance portability.}\label{tbl:example-pp}
 \end{subtable}
 \caption{A constructed example with five platforms.}\label{tbl:example}
\end{table}

We apply our performance portability metric to different subsets of our example platforms in Table~\ref{tbl:example}\subref{tbl:example-pp} to highlight interesting properties of the metric.  In the first subset, the inclusion of Platform D causes \ppm to be zero, regardless of which efficiency metric is used -- our metric cannot be used to obscure unsupported platforms, and this encourages application comparisons to consider only platforms that are supported by all applications.  This also may encourage comparisons across different meaningful subsets of platforms (\eg an application has a \ppm of X\% across CPUs, Y\% across accelerators and Z\% across both). The other subsets demonstrate that \ppm for one application can vary significantly depending on which platforms are being considered; the inclusion of Platform E in particular significantly lowers \ppm when application efficiency is used, because the harmonic mean tracks the minimum value.

Our performance portability metric is well-defined, albeit trivial, for a set of just one platform (A).  We see no reason to require two or more platforms, since such a restriction may be circumvented by different interpretations of ``platform'': it is entirely fair to consider two different products based on the same microarchitecture (\eg two processors with different core counts or frequencies) to be distinct platforms.

\subsection{Real-life Examples}

\noindent We also retroactively apply our metric to some previous application studies (including two of our own).  In order to be able to compute efficiency (either application or architectural), we are limited to studies which include some comparison to achievable performance across multiple platforms.  Note that because these studies were performed by different researchers and at different times, each study uses a different set of platforms\footnote{Configurations are available in the original papers, as specified by\\Tables~\ref{tbl:gpustream-pp}, \ref{tbl:gpustream-base} and \ref{tbl:studies}. See also the disclaimers in \S{}~\ref{sec:disclaimers}.} -- the performance portability measurements here are used to demonstrate the utility of the metric, and not to directly compare different research efforts.

\subsubsection{GPU-STREAM}

\noindent The GPU-STREAM benchmark~\cite{Deakin-GPUSTREAM} is a reimplementation of McCalpin's STREAM benchmark~\cite{STREAM} in 7 different programming models, and its authors provide performance numbers for 12 different hardware platforms.  Since STREAM is bandwidth-bound (by design), we can easily compute both architectural and application efficiency from these results: the theoretical peak performance corresponds to the peak bandwidth reported in the platform's specification sheet; and the best-known performance is the highest achieved by any GPU-STREAM implementation on the platform.

\begin{table}[t!]
 \small
 \centering
 \begin{tabular}{c|c|c}
  \multirow{2}{*}{\textbf{Implementation}} & \multicolumn{2}{c}{\textbf{\boldmath$\pp(a,p,H)$}} \\
                                           & \textbf{Arch. Eff.} & \textbf{App. Eff.} \\ \hline
  McCalpin     & 73.39\% & 96.33\% \\ \hline
  SYCL         & 55.70\% & 68.79\% \\ \hline
  RAJA         & 64.66\% & 85.68\% \\ \hline
  Kokkos       & 64.69\% & 85.66\% \\ \hline
  OpenMP (C++) & 61.01\% & 80.86\% \\ \hline
  OpenACC      & 50.02\% & 65.79\% \\ \hline
  CUDA         & 46.41\% & 64.21\% \\ \hline
  OpenCL       & 50.04\% & 65.27\% \\ \hline
 \end{tabular}
 \caption{Performance portability of GPU-STREAM 2.0, where $H$ is the subset of platforms supported by the implementation. Configurations can be found in \cite{Deakin-GPUSTREAM}, \S{} 4.}\label{tbl:gpustream-pp}
\end{table}

The results in Table~\ref{tbl:gpustream-pp} show the performance portablity for each implementation of GPU-STREAM on the subset of platforms that the implementation supports.  From this, we see three clear groupings: 1) McCalpin; 2) RAJA, Kokkos and OpenMP (C++); and 3) SYCL, OpenACC, CUDA and OpenCL.  These results demonstrate the danger in comparing performance portability computed with different platform sets -- McCalpin has the highest performance portability across the platforms that it supports (and this is useful information), but the other implementations support a greater number of platforms based on a wider variety of microarchitectures.

\begin{table*}[th]
 \small
 \centering
 \begin{tabular}{c||c|c||c|c||c|c}
  \multirow{2}{*}{\textbf{Implementation}} & \multicolumn{2}{c||}{\textbf{\boldmath$\pp(a,p,\textrm{CPUs})$}} & \multicolumn{2}{c||}{\textbf{\boldmath$\pp(a,p,\textrm{GPUs})$}} & \multicolumn{2}{c}{\textbf{\boldmath$\pp(a,p,\textrm{CPUs} \cup \textrm{GPUs})$}}\\
                                           & \textbf{Arch. Eff.} & \textbf{App. Eff.} & \textbf{Arch. Eff.} & \textbf{App. Eff.} & \textbf{Arch. Eff.} & \textbf{App. Eff.} \\ \hline
  McCalpin     & 75.66\% &  99.76\% &   0.00\% &   0.00\% &   0.00\% &   0.00\% \\ \hline
  SYCL         &  0.00\% &   0.00\% &   0.00\% &   0.00\% &   0.00\% &   0.00\% \\ \hline
  RAJA         & 58.11\% &  76.08\% &  72.92\% &  99.22\% &  63.88\% &  84.88\% \\ \hline
  Kokkos       & 57.42\% &  75.08\% &  73.23\% &  99.63\% &  63.51\% &  84.32\% \\ \hline
  OpenMP (C++) & 57.49\% &  75.37\% &  68.00\% &  92.42\% &  61.73\% &  82.10\% \\ \hline
  OpenACC      & 38.21\% &  50.00\% &  70.22\% &  95.64\% &  47.91\% &  63.46\% \\ \hline
  CUDA         &  0.00\% &   0.00\% &  72.37\% &  98.50\% &   0.00\% &   0.00\% \\ \hline
  OpenCL       & 35.28\% &  46.15\% &  73.22\% &  99.65\% &  45.84\% &  60.61\% \\ \hline
 \end{tabular}
 \caption{Performance portability of GPU-STREAM 2.0 for three platform sets. Configurations can be found in \cite{Deakin-GPUSTREAM}, \S{} 4.}\label{tbl:gpustream-base}
\end{table*}

\begin{table*}[th]
 \small
 \centering
 \begin{tabular}{c|c|c|c|c|c|l}
  \multirow{2}{*}{\textbf{Application}} & \multirow{2}{*}{\textbf{Problem}} & \multirow{2}{*}{\textbf{Model}} & \multirow{2}{*}{\textbf{Approach}} & \multicolumn{2}{|c|}{\textbf{\boldmath$\pp(a,p,H)$}} & \multicolumn{1}{c}{\multirow{2}{*}{\textbf{Configuration}}} \\
  & & & & \textbf{Arch. Eff.} & \textbf{App. Eff} & \\ \hline
     miniMD &             256k &  OpenCL & Specialization &  1.60\% &  54.03\% & \cite{Pennycook-MD}, \S{} III-C \\ \hline
     NAS-LU &          Class C &  OpenCL & Specialization &       - &  76.32\% & \cite{Pennycook-OpenCL},  \S\S{} 5.1 and 6.1\\ \hline
  D3Q19-BGK &          $128^3$ &  OpenCL &       Agnostic & 31.21\% &  91.70\% & \cite{SMS-Grid}, \S\S{} 3 and 5.1 \\ \hline
   ROTORSIM &         Cylinder &  OpenCL &       Agnostic & 29.67\% & 100.00\% & \cite{SMS-Grid}, \S\S{} 3 and  5.2\\ \hline
 CloverLeaf & $1920\times3840$ &  OpenCL &       Agnostic & 37.31\% &  90.97\% & \cite{SMS-Grid}, \S\S{} 3 and 5.3 \\ \hline
 CloverLeaf &         $3840^2$ &     OPS &            DSL & 95.96\% &        - & \cite{OPS}, \S{} V \\ \hline
        LUD &             2048 & OpenACC &    Auto-tuning &       - &  46.75\% & \cite{Sabne-OpenACC}, Table 4 and \S{} 4.1 \\ \hline
    HOTSPOT &         $4096^2$ & OpenACC &    Auto-tuning &       - &  93.06\% & \cite{Sabne-OpenACC}, Table 4 and \S{} 4.1 \\ \hline
       SpMV &             2.7m &  OpenCL &       Agnostic &  5.93\% &        - & \cite{Zhang-OpenCL}, Table 1 and \S{} 2.1 \\ \hline
 \end{tabular}
 \caption{Performance portability results for some previous application studies.}\label{tbl:studies}
\end{table*}

Table~\ref{tbl:gpustream-base} presents the same performance data relative to three different platform sets: 1) CPUs (five x86-based processors); 2) GPUs (four NVIDIA\footnote{Other names and brands may be claimed as the property of others.} GPUs); and 3) the union of the sets in 1) and 2).  No implementation of GPU-STREAM runs across all 12 of the platforms, but the nine platforms in these subsets are supported by five of the programming models.

The \ppm measurements across CPUs are notably lower than the equivalent measurements across GPUs, and this is reflected in the measurements across the union of both subsets; the fastest way for the authors of GPU-STREAM to improve the performance portability of their benchmark across \emph{both} CPUs and GPUs is to close the performance gap between GPU-STREAM and STREAM on CPUs.  Whether or not it is possible to match STREAM performance using each of these programming models on CPUs (\ie whether the programming model forces the algorithm to be structured in a way that limits performance) remains to be seen, but exposing such imbalance in platform support is important -- it allows end-users to make more informed decisions regarding application/platform selection, and encourages application developers to expend more effort in the platforms where efficiency is lowest.

\subsubsection{Other Benchmarks}

Table~\ref{tbl:studies} presents \ppm values computed using a combination of architectural and application efficiencies from a number of additional performance portability studies.  For each study, we note not only the programming model used but also the design philosophy of the developers: performing some \emph{specialization} of the source code for each platform; employing only general optimizations that are \emph{agnostic} to platform selection; using a \emph{DSL} to generate efficient binaries for each platform; and \emph{auto-tuning} a parameterized implementation to find the best fit for a platform.

There is not enough data here to draw conclusions about the effectiveness of different development approaches and programming models, but variation in the metric across studies and approaches demonstrates the importance of an objective and comparable measure of performance portability.


%% file: discussion.tex
\section{Discussion}

\noindent \ppm is a useful aggregate measure of performance and portability, but it is only meaningful when published alongside the set of platforms and the problem used to compute it.  When reading a \ppm measurement, careful attention should be paid to the platforms and problem the authors have chosen, and generalizations beyond what was recorded should be made only with great caution; as with any metric~\cite{12Ways}, it is possible to hide and misrepresent information.

The simplest (and most meaningful) way to use our metric is to compare performance portability when only one variable is changed (\eg one application executing one problem across multiple platform sets, multiple applications executing one problem across one platform set, etc).  Comparing the performance portability of different applications executing different problems on different platform sets \emph{may} be useful to draw broad conclusions regarding programming model and design philosophy choices, but any such comparisons should be made attentively (and with many more datapoints).

Although both types of performance efficiency considered in this \paper  are compatible with the calculation of \ppm, it is clear from our examples that they have different strengths.  Given multiple implementations of an application and an in-depth knowledge of the problem, application efficiency is more representative -- it conveys not only the application's performance, but also any performance penalty arising from the abstraction that it relies upon to provide portability.  However, we acknowledge that it may be difficult to compute in practice (since it requires a pre-existing best-known performance result or an accurate performance model).  Given a single implementation of an application and no pre-existing knowledge of expected performance, architectural efficiency is more representative (since the application efficiency is 100\% on every platform!), but it obscures cases where better algorithms would have provided better performance on some platforms.  The efficiencies are complementary, and presenting both is the easiest way to address their shortcomings.

%% file: summary.tex
\section{Summary}

\noindent We have proposed a novel, quantitative metric for performance portability.  This metric is simple to compute, is representative of an application's performance \emph{and} portability, and can be used to meaningfully compare the performance of different applications and high-level frameworks across the same base set of hardware platforms.

Adoption of our metric for performance portability is an important step towards productive debate, discussion and collaboration on the topic.  We anticipate further demonstrations of its utility by other studies, and intend to investigate the degree to which existing programming models and design philosophies permit or prevent high levels of performance portability in future work.
%